\documentclass{article}
\pdfoutput=1

\usepackage{microtype}
\usepackage{graphicx}
\usepackage{subfigure}
\usepackage{booktabs}
\usepackage{amsfonts}
\usepackage{amsmath}

\usepackage[obeyFinal,textsize=footnotesize]{todonotes}

\usepackage{droid}

\usepackage{hyperref}
\usepackage[cachedir=., frozencache]{minted}

\usepackage[preprint]{sysml2019}

\sysmltitlerunning{Compiling Julia to TPUs}

\newcommand{\red}[1]{}
\renewcommand{\red}[1]{{\color{red}{#1}}}

\begin{document}

\twocolumn[
\sysmltitle{Automatic Full Compilation of Julia Programs and ML Models to Cloud TPUs}

\begin{sysmlauthorlist}
\sysmlauthor{Keno Fischer}{jc}
\sysmlauthor{Elliot Saba}{jc}
\end{sysmlauthorlist}

\sysmlaffiliation{jc}{Julia Computing, Inc.}

\sysmlcorrespondingauthor{Keno Fischer}{keno@juliacomputing.com}

\sysmlkeywords{Machine Learning, Hardware Accelerators, TPUs, Julia, XLA, Compilers for Dynamic Languages}

\vskip 0.3in

\begin{abstract}
Google's Cloud TPUs are a promising new hardware architecture for machine learning workloads. They have powered many of Google's milestone machine learning achievements in recent years. Google has now made TPUs available for general use on their cloud platform and as of very recently has opened them up further to allow use by non-TensorFlow frontends. We describe a method and implementation for offloading suitable sections of Julia programs to TPUs via this new API and the Google XLA compiler. Our method is able to completely fuse the forward pass of a VGG19 model expressed as a Julia program into a single TPU executable to be offloaded to the device. Our method composes well with existing compiler-based automatic differentiation techniques on Julia code, and we are thus able to also automatically obtain the VGG19 backwards pass and similarly offload it to the TPU. Targeting TPUs using our compiler, we are able to evaluate the VGG19 forward pass on a batch of 100 images in 0.23s which compares favorably to the 52.4s required for the original model on the CPU. Our implementation is less than 1000 lines of Julia, with no TPU specific changes made to the core Julia compiler or any other Julia packages.
\end{abstract}
]

\printAffiliationsAndNotice{}
\section{Introduction}

One of the fundamental changes that has enabled the steady progress of machine learning techniques over the past several years has been the availability of vast amounts of compute power to train and optimize machine learning models. Many fundamental techniques are decades old, but only the compute power available in recent years was able to deliver sufficiently good results to be interesting for real world problems. A significant chunk of this compute power has been available on Graphics Processing Units (GPUs) whose vector compute capability, while originally intended for graphics have shown to deliver very good performance on the kind of matrix-heavy operations generally performed in machine learning models.

The real world success of these approaches and of GPUs in this space in particular has set
off a flurry of activity among hardware designers to create novel accelerators for machine learning workloads. However, while GPUs have a relatively long history of support in software systems, this generally does not extend to new, non-GPU accelerators and developing software for these systems remains a challenge.
 
In 2017, Google announced that they would make their proprietary Tensor Processing Unit (TPU) machine learning accelerator available to the public via their cloud offering. Originally, the use of TPUs was restricted to applications written using Google's TensorFlow machine learning framework. Fortunately, in September 2018, Google opened up access to TPUs via the IR of the lower level \textit{XLA} (``Accelerated Linear Algebra'') compiler. This IR is  general purpose  and is an optimizing compiler for expressing arbitrary computations of linear algebra primitives and thus provides a good foundation for targeting TPUs by non-Tensorflow users as well as for non-machine learning workloads.

In this paper, we present initial work to compile general Julia code to TPU using this interface. This approach is in contrast to the approach taken by TensorFlow \cite{tensorflow}, which does not compile Python code proper, but rather uses Python to build a computational graph, which is then compiled. It is aesthetically similar to JAX \cite{jax}, which does aim to offload computations written in Python proper by tracing and offloading high-level array operations. Crucially, however, we do not rely on tracing, instead we leverage Julia's static analysis and compilation capabilities to compile the full program, including any control flow to the device. In particular, our approach allows users to take advantage of the full expressiveness of the Julia programming language in writing their models. This includes higher-level features such as multiple dispatch, higher order functions and existing libraries such as those for differential equation solvers \cite{rackauckas2017differentialequations} and generic linear algebra routines. Since it operates on pure Julia code, it is also compatible with the Zygote.jl \cite{Zygote} automatic differentiation tool, which performs automatic differentiation as a high-level compiler pass. Putting these together, we are able to compile full machine learning models written using the Flux machine learning framework, fusing the forward and backwards model passes as well as the training loop into a single executable that is offloaded to the TPU.

The rest of this paper is organized as follows: In section \ref{section-tpu} we review the architecture of the TPU hardware. Section \ref{section-compiling-julia} reviews the workflow of the Julia compiler. In sections \ref{section-xla-embedding} and \ref{section-julia-mapping}, we present the details of embedding XLA into Julia IR and discuss some of the challenges in compiling Julia's control flow constructs to the corresponding XLA representation. We discuss challenges encountered and potential improvements to the upstream Julia compiler in section \ref{section-challenges-julia}, results in section \ref{section-results}, and give remarks on limitations of our current implementation and future work in section \ref{section-future}.

\section{TPU System Architecture}
\label{section-tpu}

\subsection{The TPU Hardware}
Google has developed three generations of TPU hardware. Both the second generation (TPUv2) and the third generation (TPUv3) are commerically available as of October 2018. Both have the ability to operate on IEEE 754 32-bit floating point numbers (\textit{float32}), as well a custom non-IEEE 16-bit floating point format (\textit{bfloat16}), that matches the bit width of IEEE 32-bit in the exponent, but trades that off for significantly reduced mantissa space. As with the previous generation, TPUv2/v3 feature a systolic array matrix multiply unit, though in this case operating using bfloat16 multiplies and float32 accumulates. At full speed, each TPUv2 core is capable of 22.5 TFLOP/s, as well as having 300GB/s bandwidth to an attached 8GB RAM of high bandwidth memory. Each TPU chip features two such cores for a total of operation speed of 45 TFLOP/s and total memory bandwidth of 600GB/s.  Additionally, TPU chips are designed to be connected in a high-performance mesh network allowing scalability to larger models and data sets. 

Google's Cloud TPUv2 offering features 4 TPUv2 chips (i.e. 8 TPUv2 cores). However, unlike most accelerator hardware, Cloud TPUs are not made available to the user directly via PCIe, rather a webservice is exposed that accepts serialized XLA IR as well as exposing lower level memory management capabilities.  This API, dubbed XRT, enables non-TensorFlow clients to generate XLA IR.  XRT went live with the deployment of TensorFlow 1.11 to Cloud TPUs on September 27th 2018.

\subsection{XLA}
\label{subsection-xla}
XLA (``Accelerated Linear Algebra'') is a partially open source compiler project by Google. It features a rich input IR for specifying multilinear algebra computations and provides backend code generation capabilities for CPUs, GPUs and TPUs. XLA's Input IR (dubbed the HLO \textit{High-Level Optimization} IR) operates on arbitrary dimensional arrays of basic data types (integers and floats of various bit widths, bfloat16s and complex numbers) or tuples thereof (but no arrays of tuples). HLO operations include basic arithmetic operations, special functions, generalized linear algebra operations, high level array operations, as well as primitives for distributed computation. XLA can perform semantic simplifications of input programs, as well as performing whole-program memory scheduling for efficient use and re-use of available memory (a very important consideration for large machine learning models). Each HLO operation has two kinds of operands:
\begin{enumerate}
    \item \textit{Static Operands} whose \textbf{values} need to be available at compile time and that configure the operation (e.g. specifying the window of a convolution operation or the summation order of a tensor contraction). Additionally some of these static operands may reference other computations that are part of the same HLO module
    \item \textit{Dynamic Operands} consisting of the aforementioned tensors (as output by other HLO operations). While the array entries of the tensor need not be available at compile time, the shape and layout of any tensor does need to be defined at compile time (i.e. there is no facility for dynamically sized dimensions).
\end{enumerate}
The mix of these operands differs depending on the instruction. E.g. the `Add` instruction takes two dynamic operands (the summands) and no static operations, while the `Constant` operation takes no dynamic operands and one static operand describing the desired constant. More complicated operations generally have a mixture of these operands. E.g. the `Map` instruction takes a static operand describing the computation to apply, as well as a variadic number of dynamic operands (though the number and shapes of these operands is of course fixed statically) describing the arrays to map over. A single dependency graph of HLO operations forms an \textit{HLO computation} and several computations form an \textit{HLO module}, though each module always has exactly one entry computation and each computation has always exactly one root operation corresponding to the returned value. HLO modules can be serialized to a protobuf-specified binary format. This is the format that is accepted by XRT.

\section{The Julia Compiler}
\label{section-compiling-julia}

In order to understand how to compile Julia code to XLA code, it is instructive to consider how the regular Julia compiler works. Julia is semantically a very dynamic language. However, in standard configuration, Julia's ultimate backend compiler is LLVM \cite{lattner2004llvm} which is a static compiler backend. The Julia compiler needs to bridge the semantic gap between the dynamic semantics of the language to the static semantics of the LLVM representation. To understand this process, we will look at four aspects of the Julia system: The dynamic semantics, the embedding of the static compiler intrinsics, interprocedural type inference and the extraction of static sub graphs. In addition we will look at the interaction of these features with macros and generated functions which will be relevant to the XLA compiler.

\subsection{Dynamic Semantics}

We begin by considering the dynamic semantics of the Julia programming language.
These are the semantics that form the mental model of the programming language
execution for the user. Additionally, these semantics are what an interpreter written
for the language would implement. Such an interpreted implementation need not be fast, merely possible.
Simpler semantics are generally better as they reduce the cognitive load on the user of the programming language. All code generated by the compiler should then operate \textit{as if} following the dynamic semantics, but ideally much faster. In Julia, the local dynamic semantics of the base language are as follows:

\begin{enumerate}
    \item All objects are (mutable or immutable), heap-allocated objects tracked by the garbage collector that contain a reference to their data type, and the actual data prescribed by that data type.
    \item A function call finds the \textbf{most specific} method \textbf{compatible} with the types of the arguments (obtained dynamically from looking at the reference in the heap-allocated objects) and transfers control there.
    \item Syntax constructs have well defined syntactic equivalents (i.e. independent of the runtime types of values of any variables) as sequences of function calls and gotos; gotos operate only on booleans. \footnote{Additional semantics include exception mechanisms, and a few corner case features, but they are not relevant to this work}
\end{enumerate}

Additionally, there are facilities for modifying the top level state (e.g. method tables and type definitions are available) at global scope, but it is semantically valid to ignore these in local semantics.

Assuming an implementation of the \textbf{most specific} and \textbf{compatible} predicates (which are highly non-trivial to implement - see \cite{juliasubtyping} - particularly in a performance-oriented manner, but are generally intuitively understandable by users), we can obtain functioning - if slow - implementations of these semantics with no more than a hundred lines of code. The relative minimality of the dynamic semantics reduces the effort required for the compiler implementation and allows work to be shared between different backends. 

\subsection{Static Compiler embedding}

Notably absent in our discussion of the dynamic semantics are any computational intrinsics. So far, we have defined what it means to call functions, but our functions can't actually do anything other than calling other functions (and performing control flow). There is a straightforward way to remedy this: We can add opaque functions to our method table that perform some specified computation (e.g. we might add a method 
\texttt{add\_int} that, given two values with \texttt{Integer} type tags, performs an integer addition on their payload bytes and returns a newly allocated result value whose payload is the result). Since we eventually compile to LLVM, the best way to pick these intrinsics is to match LLVM's intrinsics (e.g. our \texttt{add\_int} intrinsics would correspond essentially 1-to-1 to the semantics of the LLVM \texttt{add} intrinsic). Note that doing so also provides a straightforward way to implement these intrinsics: When the method is called, we look at the argument types and then (at the time of the invocation of the intrinsic) call out to the compiler to generate a specialized method that operates on boxed representations of those types and provides back a newly-allocated boxed representation of the result. This is of course again horribly slow, but very straightforward to do. It also suggests a way out of our performance trap: If we only knew ahead of time what operations were going to be executed, we could chain the intrinsics together and compile them all at once. Additionally, doing so would allow us to eliminate the expensive round trip to the heap representation and feed the input of one intrinsic directly into the output of the next.

\subsection{Interprocedural type inference}
Having formulated the question in the previous paragraph, the answer is apparent: Given some set of starting type information, perform dataflow type inference to determine, for as many subsequent operations as possible, the data types of the input values. The process is conceptually straightforward. For every intrinsic, we associate a corresponding ``transfer function'' that maps input types to output types. We then iterate the process of propagating types to convergence. Of course actually doing the latter, in the presence of arbitrary control flow and recursion while guaranteeing termination is again a non-trivial endeavor, but at least locally the process is simple. It is worth noting that unlike type inference in static languages, we do not aim for completeness. Indeed, it is impossible to define perfectly precise transfer functions for many intrinsics. Of importance to the current work is that the Julia inference lattice is significantly finer than its type lattice, allowing increased inference precision crucial for the success of this work. 

\subsection{Static sub-regions}

Having implemented type inference, we now have a performant way to implement the semantics described in the first two sections. Whenever the interpreter performs a function call, it uses the dynamic type information to perform type inference to discover and infer the largest possible static sub-region of code reachable from this entry point (recall that determining call targets requires type information). We can then hand this sub-region to the compiler and have it generate an efficient version of the entirety of the statically reachable sub-region \footnote{For performance we allow these regions to have multiple entry points, and heuristically use less than the maximum amount of available information to allow such compilations to be reused.}. Whenever the compiler encounters a call for which type inference was unable to determine the call target, it emits a call back into the runtime system (making sure to obey the dynamic semantics at the boundary) to begin the process anew.  As such, execution generally proceeds as a chain of relatively large static sub regions, chained together by the runtime system. It is only on the boundaries between these regions that the dynamic semantics are materialized (i.e. values moved to the heap, dynamic method selection performed etc.). The performance of this scheme depends heavily on the size of these static sub-regions, which is highly sensitive to both the quality of the type inference implementation and the semantics of the language. In Julia, these static subregions can easily encompass thousands of functions covering tens of thousands of source lines, without once re-entering the runtime system for method selection.



\section{The XLA embedding}
\label{section-xla-embedding}

To compile to XLA instead of LLVM, we apply the exact strategy as outlined in the previous section. In fact, we can re-use most of the compiler itself (in particular all of type inference and all mid-level optimization passes). However, let us first follow the steps outlined in the previous section and define our dynamic semantics and static embedding.

\subsection{Tensor representation}
Owing to its heritage as a teaching and research language for linear algebra, Julia has a very rich hierarchy of array abstractions. Julia's standard library arrays are mutable and parameterized over type and dimension. Additionally, the StaticArrays.jl \cite{StaticArrays.jl} package provides immutable arrays parameterized on element type and shape. As a result, the notion of shaped, N-dimensional immutable tensors is not foreign to Julia code and most existing, generic code is able to handle it without problem. We thus embed XLA values by defining a runtime structure corresponding to immutable, shaped, N-dimensional tensors backed by handles to remote, XRT-managed, memory (figure \ref{listing:XRTarray}).

\addtocounter{footnote}{-1}
\begin{listing}[h]
\begin{minted}[
frame=lines,
framesep=2mm,
baselinestretch=1.2,
fontsize=\footnotesize,
fontfamily=fdm,
xleftmargin=6pt,
linenos
]{Julia}
const AA{T, N} = AbstractArray{T, N}
struct XRTArray{T, Shp, N} <: AA{T, N}
  storage::XRTAllocation
  # XRTArrays are constructable by
  # conversion from regular arrays
  function XRTArray(
        a::Array{T, N}) where {T, N}
    new{T, size(A), N}(transfer(a))
  end
  # XRTArrays are constructable from a
  # remote memory allocation if
  # (T, Dims, N) are specified
  function XRTArray{T, Dims, N}(
      a::XRTAllocation) where {T, Dims, N}
    new{T, Dims, N}(a)
  end
end
\end{minted}
\caption{The definition of an XRTArray\protect\footnotemark. T is the element type of the array, \texttt{Shp} is a tuple of integers describing the shape of the tensor, \texttt{N} is always equal to \texttt{length(Shp)}, which is enforced by the constructor and is required because Julia does not currently allow computation in the subtype relation. The \texttt{AA} alias is defined for brevity of notation only.}
\label{listing:XRTarray}
\end{listing}

\footnotetext{This presentation is slightly simplified for this presentation. The real definition includes facilities for caching data locally in order to allow defering and batching transfer to the remote device.}

These handles play the same role as our heap allocated, boxed values when compiling to LLVM. On the boundary of static subregions, these values need to be materialized: we inspect the type of the remote allocation, noting the type and shape and then generate an XLA function that performs the desired intrinsic on the input, generating another XRTArray (backed by remote storage) as the output. These dynamic semantics are very slow, requiring at least one network roundtrip for every executed operation \footnote{Because of the coarse-grained nature of HLO operations, the resulting performance is not horrible and is relatively similar to how GPU kernels are traditionally launched. However, the TPU is significantly faster than a traditional GPU, the remote launch times are longer, and the HLO mapping from Julia is written with the expectation of being fused together (e.g. broadcasting constants to the size of an array with the expectation that the broadcasted intermediate value is never materialized)}. Once we have successfully defined this representation, we are done in terms of semantics (on top of the base julia semantics, but largely replacing the LLVM-derived semantics). We can then re-use the existing Julia compiler to give us the largest possible static sub-regions, which will then correspond to an embedding of XLA into Julia.

\subsection{Operation representation}

\subsubsection{Separating static and dynamic operands}
\label{subsection-operands}
As discussed in section \ref{subsection-xla}, HLO operands are partitioned into static and dynamic operands. Suppose we have an example  XLA operation `Foo` taking one static operand (e.g. a single integer) and two dynamic operands. We would declare this embedding as follows:

\begin{minted}[
frame=lines,
framesep=2mm,
baselinestretch=1.2,
fontsize=\footnotesize,
fontfamily=fdm,
xleftmargin=6pt,
linenos
]{Julia}
struct HloFoo <: HloOp{:foo}
  static_operand::Int
end

function (hlo::HloFoo)(dop1::XRTArray,
                       dop2::XRTArray)
    execute(hlo, dynamic_op1, dynamic_op2)
end
\end{minted}

In this example, the `execute` function implements the dynamic semantics of running the operation on the remote device.  The \texttt{function (hlo::HloFoo)(...)} syntax denotes call-operator overloading.  This therefore means that a call to \texttt{HloFoo(1)} will construct and return a callabale object that, when called on two XRTArrays, will remotely execute the `Foo` HLO operation with static operand `1` and the dynamic operands corresponding to the two arrays. This separation is not strictly necessary, but does have the useful property that embedding into Julia IR is easy to understand:

\begin{listing}[h]
\begin{minted}[
frame=lines,
framesep=2mm,
fontsize=\footnotesize,
fontfamily=fdm,
xleftmargin=6pt,
linenos
]{Julia}
# An HLO operand that generates a random
# uniform random number of the specificed
# shape and element type:
struct HloRng <: HloOp{:rng}
    Type
    Shape
end

"""A function that adds random numbers to
   each entry of a 1000x1000 matrix"""
@eval function add_rand_1000x1000(
    A::XRTArray{Float32, (1000, 1000), 2}
    random = $(HloRng(Float32,
               (1000, 1000)))()
    result = $(HloAdd())(random, A)
    return result
end
\end{minted}
\caption{
\label{listing:manual-embedding}
A manually constructed XLA embedding. \texttt{HloRng} is slightly simplified from the actual operation for clarity of presentation. The boilerplate definition calling execute is omitted.}
\end{listing}

In the example in listing \ref{listing:manual-embedding}, we have spliced the HLO operands
(including the static operands) right into the AST.  This yields a mapping to XLA that is extremely straightforward (go through every statement, obtain the static operands from the spliced instruction specification, and the dynamic shapes from type inference and generate the corresponding XLA code). Of course, we don't generally splice these instructions in manually, however the manual example illustrates why separating the static operands is useful and illustrates the conditions for a successful offload to XLA. IR is completely offloadable if, after all relevant Julia level optimizations:
\begin{enumerate}
    \item Every statement is a call to a spliced value whose type is a subtype of `HloOp`; and
    \item Every operand type is fully inferred and a subtype of `XRTArray`
\end{enumerate}
IR that satisfies these conditions is trivially transformable into XLA IR. We will loosen these conditions slightly and also make the embedding more general in subsequent sections,
but these two conditions capture the basic idea. If we can convince the Julia compiler to generate IR of this form, generating equivalent XLA IR becomes trivial (a simple rewriting from one representation to another).

\subsubsection{Representation of functions with higher-order callbacks}

There is one additional detail we have so far ignored: Several HLO operations are higher level functions and themselves take a computation to apply as a static operand. 
At this point, we generalize the embedding slightly and allow the callback to be specified as an arbitrary julia function (the dynamic semantics being application of the julia function in a manner consistent with XLA's semantics for the higher order function). \footnote{The HLO IR specification does contain provisions for calling arbitrary opaque functions via the `CustomCall` operation, but that facility is not available on TPUs (as there is no non-XLA method to generate code for the TPU)}. However, in practice the called functions tend to be very simple and generally representable in XLA, so we may obtain HLO for the called function simply by recursively invoking the compiler.

As a further complication, julia function objects can contain non-trivial data (e.g. values captured by a closure) and in particular, these objects can contain embedded HLO values. Semantically, these values are available to the interior computation, unaffected by the semantics of the operation (e.g. if the operation maps over all elements of an array, captured values are not mapped over, but instead provided to the kernel as is). Currently HLO does not allow such captured values to be passed directly to the kernel, though this feature is planned for at least the \textit{Map} operation. In preparation for this feature, we split called computations between static and dynamic operands. We make the \textit{type} of the function a static operand and the \textit{value} of the function a dynamic operand. For any simple function (those that do not have any captured data) these two are equivalent. Because of the lack of HLO support, we currently require any captured \textit{value} to be inferable as a constant and thus statically available (which we then materialize as a constant in the called computation).

\subsection{Shape Transfer Functions}
\label{shape-transfer-functions}




One additional consideration is the need for type inference to be able to statically understand the shape of each HLO operation (i.e. how the output shape depends on the input shapes). In essence, we need the equivalent of the type \textit{transfer functions} we had for LLVM IR. Luckily, since all HLO operations are precisely inferable over the type lattice (as opposed to the inference lattice) there is no need to add these transfer functions to
the compiler. Instead, we can define our execute method as:
\begin{minted}[
fontsize=\footnotesize,
fontfamily=fdm,
xleftmargin=6pt,
linenos
]{Julia}
execute(op::HloOp, args::XRTArray...) =
 _execute(op, args...)::shape_infer(op,
   map(typeof, args)...)
\end{minted}

The :: operator is the typeassert operator and has the dynamic
semantics of throwing a type mismatch error if the type of value on the
LHS of the operator is not a subtype of the type specificied on the RHS.
In this case, the RHS is itself a function call that computes a type
(types are usable as values in Julia and can be computed over). Moreover,
even if type inference cannot determine the type of the value on the left
hand side, as long as it can statically determine the value of the type on
the RHS, it can from then on assume that the type of the value on the LHS
is a subtype of the type on the righthand side. In this way, we can specify our shape transfer
functions in plain Julia and lift them into the type domain where type inference is able to make use of them.


\section{Mapping Julia semantics to XLA}
\label{section-julia-mapping}

We now have the ability to compile Julia programs to XLA, as long as those programs are written in terms of XLA primitives. Julia programs not, however, written in terms of arcane HLO operations; they are written in terms of the functions and abstractions provided by Julia's base library. Luckily Julia's use of multiple dispatch makes it easy to express how the standard library abstractions are implemented in terms of HLO operations. A few simple examples of this are shown below:

\begin{minted}[
frame=lines,
framesep=2mm,
fontsize=\footnotesize,
fontfamily=fdm,
xleftmargin=6pt,
linenos
]{Julia}
# Matrix-Matrix and Matrix-Vector product
function Base.:*(A::XRTMatrix,
    B::Union{XRTMatrix, XRTArray})
  ddots = DimNums((1,), (0,), (), ())
  HloDot(ddots)(A, B)
end
Base.transpose(A::XRTArray) =
    HloTranspose((1,0))(A)
# Scalar addition
Base.:+(A::XRTArray{T, (), 0},
        B::XRTArray{T, (), 0})
    where {T<:XLAScalar} =
 GenericHloOp{:add}(T, ())(A, B)

\end{minted}

In addition to these simple operations, we also provide implementations of
the higher level array abstractions, in particular, \texttt{mapreduce} and \texttt{broadcast}. The implementation of \texttt{broadcast} in terms of HLO operations is about 20 lines of code and omitted for space, but the implementation of `mapreduce` is simply:

\begin{minted}[
frame=lines,
framesep=2mm,
fontsize=\footnotesize,
fontfamily=fdm,
xleftmargin=6pt,
linenos
]{Julia}
dims_tuple(A, ::Colon) = tuple(
    (0:ndims(A)-1)...)
dims_tuple(A, t::Tuple) = t
dims_tuple(A, n::Int) = (n,)
function Base.mapreduce(f, op, A::XRTArray;
        dims=:)
    HloReduce(op, dims_tuple(A, dims))(
        HloMap(f)(A),
        XRTArray(
            Base.mapreduce_empty(f, op,
                eltype(A)))
    )
end
\end{minted}

In this, we can see the utility of allowing arbitrary Julia functions as static computation operands. Thanks to Julia's reliance on generic abstractions, it suffices to specify very few of these definitions in order to cover a large surface of APIs. In particular, from the mapreduce definition above we automatically get any reduction defined in base e.g. \texttt{sum} and \texttt{prod}. In fact, obtaining sufficient coverage to compile both the forward and the backwards passes of the VGG19 computer vision model \cite{vgg19}, requires less than 200 lines of definitions.

\subsection{Structure mapping}

We make one additional identification. Any tuple or immutable structure present in the embedded IR ges mapped to an XLA tuple. I.e. the julia value $1+2im$ (a complex number represented by a struct of two integers), would be mapped to the XLA tuple $(s64[], s64[])$. We preserve the struct type of these in the Julia embedding of XLA IR, but naturally XLA has no notion of julia types so they get translated to proper tuples during the final translation step. Similarly, (julia) \textit{Tuple} constructors (and constructors for immutable structs) become tuple construction on the XLA side. Tuple references (field references for immutable structs) become tuple references on the XLA side.

\subsection{Handling control flow}
\label{section-control-flow}

One additional complication we have not yet discussed is the semantic mismatch between the imperative control flow offered by Julia and the functional control flow offered by XLA. To handle \texttt{if}/\texttt{else} blocks, we look at $\phi$ nodes in the julia compiler's SSA IR. We then take these $\phi$ nodes as the result value of XLA's functional control flow (if there are multiple $\phi$ nodes at the same merge point we form a tuple of these $\phi$ nodes). The condition that originally caused the divergence becomes the condition of the functional control flow. Any computation in between is outlined and attached as a called function. Loop identification works similarly. We identify strongly connected regions of the control flow graph, and outline these as the loop body. We combine values that have uses outside the strongly connected region as well as any loop carried $\phi$s into the iteration state of the functional while loop (since most loops can be empty we often have a $\phi$ node that corresponds to uses of loop values outside the loop).

\section{Inference puzzles}

Performing the analysis required by this technique puts significant burden on Julia's type inference to infer code very precisely. In order to be offloadable to XLA, type inference needs to be able to figure out every static operand to every HLO instruction as well as the shapes of every dynamic operand. In addition, it needs to be able to constant fold or otherwise eliminate all utility computations that are not expressed in terms of HLO operations. Julia's type inference includes a number of heuristics that are supposed to prevent excessive time spent in inference when there would be little runtime benefit. However, for offloading to XLA, these heuristics are mistuned.  In addition, parts of Julia's type inference are not designed to handle such a large number of constants as are required to generate a sufficiently high quality XLA embedding. For example, inference results are cached based on the precise, concrete signatures of the operands. In our case, the shapes of operands are part of the signature, thus forcing inference to re-infer every function for every new signature. This could be significantly improved if instead of caching concrete signatures, type inference was able to infer transfer functions for inference results, thus allowing it to re-use work, even if the exact shape values differ. Fortunately, at the moment these concerns are largely academic since the time spent in XLA itself significantly dwarfs the time taken by type inference to infer the input IR.

The results in this paper were obtained with a custom version of Julia that disabled several limiting heuristics and fixed a number of bugs leading to suboptimal inference. We are in the process of contributing these improvements back to Julia (for the limiting heuristics as options to the compiler interface to request they be disabled for a particular invocation). Other than that, no TPU or XLA specific changes had to be made to Julia and all functionality described in this paper lives entirely within a self-contained Julia package.

\label{section-challenges-julia}

\section{Results}
\label{section-results}

The method described in this paper heavily relies on the Julia middle end compiler to determine sufficiently precise information (in particular the values of static operands and the shapes of dynamic operands), in sufficiently large subregions of the program to amortize any launch overhead. In this section, we demonstrate that the Julia compiler (with the modifications described in section \ref{section-challenges-julia}) is indeed precise enough to make this method applicable to program of practical interest.

\subsection{Two simple examples}
\label{simple}

Before moving on to more complicated examples, let us consider two simple examples that are subproblems of the full VGG19 example:

\begin{minted}[
frame=lines,
fontsize=\footnotesize,
fontfamily=fdm,
xleftmargin=6pt,
linenos
]{Julia}
dense(W, x, b) = (W * x) .+ b
softmax(xs) = exp.(xs) ./ sum(exp.(xs))
\end{minted}

Our implementation provides introspection macros inspired by those provided
by base Julia itself. In particular, in base Julia, \texttt{@code\_lowered} provides the state of the code after parsing, macro expansion and lowering, \texttt{@code\_typed} provides the state of the code after type inference and mid-level optimizations, and \texttt{@code\_llvm} provides the generated llvm code. Analogously, We provide \texttt{@code\_typed\_xla} for showing the typed IR after our enhanced optimization passes (described in section \ref{section-challenges-julia}) as well as \texttt{@code\_xla} for printing the resulting XLA IR in its native textual representation. For the \textit{dense} example, we show steps along the full pipeline, in particular after inference (listing \ref{listing-dense-after-inference}), inlining and optimizations (listing \ref{listing-dense-after-opt}) as well as the final XLA IR (listing \ref{listing-dense-final}),

\begin{listing}[ht]
\begin{minted}[
frame=lines,
fontsize=\footnotesize,
fontfamily=fdm,
xleftmargin=6pt,
linenos
]{Julia}
@code_typed_xla opt=false dense(W, x, b)
 %1 = Base.Broadcast.materialize::Const(
        materialize, false)
 %2 = Base.Broadcast.broadcasted::Const(
        broadcasted, false)
 %3 = (W * x)::XRTArray{Float32,(10,),1}
 %4 = (%2)(Main.:+, %3, b)::Broadcasted{
        XRTArrayStyle{1},Nothing,typeof(+),
        Tuple{XRTArray{Float32,(10,),1},
        XRTArray{Float32,(10,),1}}}
 %5 = (%1)(%4)::XRTArray{Float32,(10,),1}
      return %5
\end{minted}
\caption{\label{listing-dense-after-inference}State of the IR after type inference. Note that the system was able to determine both the shape and element type of the matrix multiply and the broadcast operations. In particular, for the matrix multiply, it invoked the shape transfer function we defined in section \ref{shape-transfer-functions}. The output was edited slightly to remove line number information and add line breaks.}
\end{listing}

\begin{listing}[ht]
\begin{minted}[
frame=lines,
fontsize=\footnotesize,
fontfamily=fdm,
xleftmargin=6pt,
linenos
]{Julia}
@code_typed_xla opt=true dense(W, x, b)
 %1 = invoke HloDot(XLA.DimNums{1,1,0,0}(
    (1,), (0,), (), ()))(_2, _3)::XRTArray
        Float32,(10,),1}
 %2 = invoke HloMap{typeof(+)}(+)(
    %1, _4)::XRTArray{Float32,(10,),1}
      return %2
\end{minted}
\caption{\label{listing-dense-after-opt}State of the IR after inlining and optimization. Note that the representation follows the constraints specified at the end of section \ref{section-xla-embedding}: Every instruction is a fully specified (constant value of HloOp type) and every operand shape is fully inferred (\texttt{\_i} by virtue of being arguments, \texttt{\%1} by virtue of the correct inference of the return type of \texttt{HloDot}). This is possible because the Julia compiler aggressively optimized out all non-essential objects and abstraction (e.g. the \textit{Broadcasted} object we saw in the previous listing).}
\end{listing}

\begin{listing}[ht]
\begin{minted}[
frame=lines,
fontsize=\footnotesize,
fontfamily=fdm,
xleftmargin=6pt,
linenos
]{Julia}
@code_xla opt=true dense(W, x, b)
 c1 {
   c1p0 = f32[] parameter(0)
   c1p1 = f32[] parameter(1)
   ROOT c1a2 = f32[] add(c1p0, c1p1)
 }

 ENTRY dense {
   c0p0 = f32[10,10]{0,1} parameter(0)
   c0p1 = f32[10]{0} parameter(1)
   c0d3 = f32[10]{0} dot(c0p0, c0p1),
     lhs_contracting_dims={1},
     rhs_contracting_dims={0}
   c0p2 = f32[10]{0} parameter(2)
   ROOT c0m4 = f32[10]{0} map(c0d3,
     c0p2), dimensions={0}, to_apply=c1
 }
\end{minted}
\caption{\label{listing-dense-final}The final XLA IR, ready for XLA's high level optimization and eventual generation of TPU machine code. Notice that the $+$ being broadcasted turned into a separate computation computing scalar addition (We saw the corresponding scalar $+$ definition in section \ref{section-julia-mapping}. The listing was edited to introduce line breaks and shorten autogenerated variable names.}
\end{listing}

\begin{listing}[ht]
\begin{minted}[
frame=lines,
fontsize=\footnotesize,
fontfamily=fdm,
xleftmargin=6pt,
linenos
]{Julia}
@code_xla opt=true dense(x)
 c1 {
   c1p0 = f32[] parameter(0)
   ROOT c1e1 = f32[] exponential(c1p0)
 }
 c2 {
   ROOT c2p0 = f32[] parameter(0)
 }
 c3 {
   c3p0 = f32[] parameter(0)
   c3p1 = f32[] parameter(1)
   ROOT c3a2 = f32[] add(c3p0, c3p1)
 }
 c4 {
   c4p0 = f32[] parameter(0)
   c4p1 = f32[] parameter(1)
   c4t2 = (f32[], f32[]) tuple(c4p0, c4p1)
   c4gte3 = f32[] get-tuple-element(c4t2)
     index=0
   c4e5 = f32[] exponential(c4gte3)
   c4gte4 = f32[] get-tuple-element(c4t2),
     index=1
   ROOT c4d6 = f32[] divide(c4e5, c4gte4)
 }
 ENTRY softmax {
   c0p0 = f32[10]{0} parameter(0)
   c0m1 = f32[10]{0} map(c0p0),
     dimensions={0}, to_apply=c1
   c0m2 = f32[10]{0} map(c0p0),
     dimensions={0}, to_apply=c2
   c0c3 = f32[] constant(0)
   c0r4 = f32[] reduce(c0m2, c0c3),
     dimensions={0}, to_apply=c4
   c0b5 = f32[10]{0} broadcast(c0r4),
     dimensions={}
   ROOT c0m6 = f32[10]{0} map(c0p0, c0b5),
     dimensions={0}, to_apply=c4
 }
\end{minted}
\caption{\label{listing-softmax-final}The final XLA IR for the softmax example. See section \ref{simple}.}
\end{listing}

For softmax we show the final XLA IR only (figure \ref{listing-softmax-final}) for space efficiency. There are several interesting aspects to this listing that are worthy of mention. Let us first consider how the \texttt{sum} invocation is represented in the XLA IR. The Julia standard library contains the following definition of sum:
\begin{minted}[
frame=lines,
fontsize=\footnotesize,
fontfamily=fdm,
xleftmargin=6pt,
linenos
]{Julia}
add_sum(a::T, b::T) where {T} = T
sum(f, a) = mapreduce(f, add_sum, a)
sum(a) = sum(identity, a)
\end{minted}
We saw how to translate \texttt{mapreduce} operations to HLO in section \ref{section-julia-mapping} and indeed, we can see this play out in the final IR for softmax. The \textit{c0m2} operation corresponds to the map of identity \textit{identity} (represented as computation \textit{c2}) over the array, while the reduction \textit{c0r4} corresponds to the reduction using `+` (represented as computation \textit{c3}). An additional interesting feature of the softmax IR is computation \textit{c4}. It is generated from \textit{syntactic broadcast fusion} \cite{moredots}, a Julia feature for more efficient broadcasts in the absence of an optimizing array compiler (and is thus likely not beneficial for the XLA backend, but not hurtful either). In particular, through a combination of parser and library support, the definition of \texttt{softmax} is essentially rewritten to

\begin{minted}[]{Julia}
softmax(xs) = broadcast(
    (args...)->exp(args[1])/args[2],
    xs, sum(exp.(xs)))
\end{minted}

Thus \textit{c4} corresponds to the closure passed to broadcast (fusing the exponential and the division). The additional packing and unpacking of a tuple corresponds to constructing the \texttt{args} argument tuple for the closure \footnote{These operations are inserted by the final compiler step, translating the XLA embedding into proper XLA IR. Had they been inserted earlier julia-level destructuring passes would have optimized away such a construct. Similar passes exist on the XLA level and will optimize these operations away before final code generation.}. As is evident from this IR, we rely on XLA to perform high level simplification passes. We could attempt to perform such optimizations on the embedding instead, but performing them after translation to XLA allows a cleaner separation of concerns, as well as sharing work with other frontends. At that point, no Julia specific information is required anymore for optimization (the same principle applies to LLVM - the Julia compiler generally does not implement or even if implemented does not enable optimizations if they don't require Julia-specific information and instead defers such optimizations to LLVM).

\subsection{VGG19 forward pass}

Our first, more complex example is the full VGG19 forward pass. We use the implementation of VGG19 as found in the Metalhead package \cite{Metalhead.jl}, which leverages the Flux \cite{Flux.jl} framework to translate the familiar machine learning layers (convolutional layer, dense layer) into linear algebra operations. However, importantly each layer in the Flux framework is just a regular function that in turn calls regular linear algebra operations. As such, machine learning models expressed in Flux, including VGG19, are just simply regular Julia functions and thus amenable to the methods described in this paper. Our compiler is able to fully infer, offload and fuse the entire forward pass of VGG19. After Julia-level optimizations, the final IR for the top level function contains 181 instructions (each an HloOp with properly inferred constant static parameters and properly shape inferred dynamic paramters. The total number HLO operands in the entry level computation is 183 (two extra for the parameter instructions which are implicit in the embedding) and 361 total over 29 computations\footnote{However, some of these computations are identical to each other because of the regular structure of VGG19. For simplicity our compiler currently does cache and re-use generated XLA IR, though that is a planned enhancement.}. The count of instructions is summarized \ref{tab:all_instructions} (a more detailed breakdown can be found in the appendix). Since we are able to offload the entire forward pass computation, the Julia is not involved at any step of the evaluation and can thus simultaneously perform other tasks (e.g. data preparation for the next batch). Additionally, the performance of the resulting code is limited only by the quality of the code generated by XLA, not by frontend considerations (we perform a performance evaluation in section \ref{sec:perf}). We validated correctness of the generated XLA code by evaluating the VGG19 model on images from the ImageNet validation set and validating that the obtained results match the results obtained from vanilla Metalhead (up to minor floating point rounding differences that generally don't affect the prediction).

\begin{figure}
    \centering
    \begin{tabular}{ r  r | c c}
    VGG19 & & Entry & Total \\ \hline
    Forward  & Unopt & 183 & 361 \\
             & Opt & 130 & 242 \\ \hline
    Backward & Unopt & 577 & 2775 \\
             & Opt & 362 & 925
    \end{tabular}
    \caption{Summary of XLA instruction generated by the Metalhead.jl VGG19 forward pass and backwards pass after compilation to XLA. Both unoptimized (after the Julia frontend) and optimized counts (after an XLA optimization pipeline similar to that used by the CPU backend, but without HLO fusion) are shown. For each, the count is further broken down into instructions in the entry (top-level) computation and instruction counts in all computations. An expanded version of this table with breakdowns by instructions is given in the appendix in figure \ref{tab:all_instructions}}
    \label{tab:instructions_summary}
\end{figure}

\subsection{VGG19 backward pass}

To obtain the backwards pass, we make use of the Zygote.jl compiler-based AD framework \cite{Zygote}. Zygote operates on Julia code and its output is again a Julia function (suitable for reintroduction into Zygote to obtain higher order derivatives, but also suitable for compilation to TPUs).

In particular, the example we are looking at is:
\begin{minted}[
fontsize=\footnotesize,
fontfamily=fdm,
xleftmargin=6pt,
]{Julia}
using Zygote
backwards(m::VGG19, x) = derivative(
    m -> sum(m(x)), m)
\end{minted}
i.e. the derivative with respect to the model at the current value of the model and a particular training example (or a batch of training examples). We use \textit{sum} as a simple stand in for the loss function. Fortuitously, but not entirely coincidentally, the type inference modifications we describe in section \ref{section-challenges-julia} also improve the precision of type inference to be able to infer through all of the VGG19 backwards pass. As for the forward pass, the total optimized and unoptimized instruction counts are shown in figure \ref{tab:instructions_summary}. The backwards pass generates significantly more XLA instructions than the forward pass. One of the biggest contributors to the instruction bloat is Zygote's mixed mode broadcast fusion, which computes both the forward pass and the backwards pass in one \textit{map} kernel. Because XLA currently does not support multiple outputs from one map instruction, the function body gets duplicated across multiple map instructions, which XLA's DCE then needs to clean up. In general, our compilation process stresses XLA's handling of the \textit{map} instruction, because of the prevalance of calls to julia's \textit{map} and \textit{broadcast} functions in generic code. We are in the process of improving XLA's handling of \textit{map} to inline mapped computations providing XLA backends with a form of IR more similar to that generated by other frontends.

\subsection{Evaluation on TPUs}
\label{sec:perf}

In this section, we present preliminary performance results of the code generated via the method in this paper. Note that because we were able to fully offload the function of interest, we expect future performance improvements to be the result of improvements to XLA's high level and backend optimizations, as opposed to modifications to the method in this paper. We note that the XLA developers have not yet had a chance to implement any improvements as a result of our work and we thus expect all XLA performance results to improve in the future. Our results are shown in figure \ref{figure:perf-results}. 

\begin{figure}
    \centering
    \begin{tabular}{ r || c | c | c | c | c }
      N= & 1 & 10 & 100 \\ \hline \hline
      Flux CPU & 0.79s & 6.67s & 52.4s \\ \hline
      PyTorch CPU & 1.16s & 9.55s & 93.0s \\ \hline FluXLA CPU & 12.06s & 64.8s & $>600$s  \\ \hline
      FluXLA TPU (total) & 0.86s & 0.74s & 0.93s  \\ \hline
      FluXLA TPU (compute) & 0.02s & 0.04s & 0.23s \\ \hline

    \end{tabular}
    \caption{\label{figure:perf-results}Timings for the VGG19 forward pass for varying batch sizes. Flux CPU is Flux master/Julia master without the XLA compiler. PyTorch CPU is the equivalent model in pytorch on the same CPU. FluXLA CPU is our work against an xrt implementation running on the CPU, FluXLA TPU (total) is end-to-end time as reported by the client (including kernel launch overhead and data transfer back from Google Cloud over the internet - note that as a result of the additional network transfer this measurement had significant variability), FluXLA TPU (compute) is the total compute time on TPUs as reported by the cloud profiler (unlike the total time measurement, this measurement was very stable). All CPU measurements on \textit{Intel(R) Xeon(R) Silver 4114 CPU @ 2.20GHz} CPUs supporting AVX512. Up to 20 cores were available and CPU benchmarks were not restricted to a single core (though in practice not all CPU benchmarks actually used the available parallism). TPU benchmarks were restricted to a single TPU core.
    All Timings are minimum of 4 runs (except FluXLA CPU for N=100 which failed to finish a single run within 10 minutes).}
    
    \vspace{-8pt}
\end{figure}

\section{Limitations and Future Work}

The obtained results are very promising and show the feasibility and generality of this approach to mapping Julia code to XLA and thus compiling to TPUs. However, significant challenges remain. For one, the current compilation model is very ``all or nothing''. The largest subregions considered for offloading are those that consist (after optimizations), entirely of XLA operations. As shown above, this is not a bad situation and is sufficient for applications of real world interest. However, we can do better. Right now, we terminate such static regions, even if intervening instructions do not have a data dependency on the intervening region. A better approach would be to use the compiler to automatically separate functions into offloadable and non-offloable parts and insert infeed/outfeed operations for any data dependencies. This will also require extending the dynamic semantics with synchronization intrinsics.
An additional area of shortcoming is our coverage of XLA's distributed computing primitives. This is partly due to lack of access to hardware where these primitives can be used effectively. It will be interesting to integrate these primitives with the rest of Julia's distributed computing capabilities.

Additionally, the debuggability of a failure to infer and offload is poor. While seasoned Julia users are able to use the existing tools to understand why the compiler was unable to make a given determination, this often seems like a black art to less experienced users. It is desirable to extend these tools with convenient, human-readable explanations of a type inference failure such that users may take appropriate remedies.

Lastly, the element types of XRTArrays are restricted to those supported by XLA. However, part of the appeal of Julia is that most code is generic over datatypes. XLA's limitation restricting element types can be to some extent overcome by performing AoS $\rightarrow$ SoA transformations (which is possible in Julia by using the \texttt{StructsOfArrays} package \cite{StructsOfArrays.jl}).

\label{section-future}

\section{Conclusion}

In this paper, we discussed how to compile Julia code to XLA IR, thus enabling offload to TPU devices. The described implementation re-uses significant parts of the existing Julia compiler and is thus less than 1000 lines of code, but is nevertheless able to compile both the forward and the backward pass (and the fusion thereof, including the training loop) of models of practical interest such as VGG19 into a single XLA kernel. We have also demonstrated how Julia's multiple dispatch semantics aid in the specification of this transformation. This work suggests that it is possible to not only compile a number of ML models written in Julia to TPUs, but also more general non-ML Julia code (as long as such code is also dominated by linear algebra operations). We hope that this facility may hasten the exploration of non-ML problem areas for which TPUs may be useful.

\section*{Acknowledgements}

Our work heavily leverages Julia type inference capabilities, which were recently significantly enhanced by Jarrett Revels, Jameson Nash and Jeff Bezanson in support of the Cassette.jl dynamic compiler framework \cite{Cassette.jl}. We are indebted to Mike Innes for his work on Flux.jl and Zygote.jl, without which we would not have been able to show the applicability of our method to a model of real world interest. Matt Bauman kindly provided guidance on properly implementing Julia's broadcast semantics against the XLA backend. More generally, we thank the Julia community for their attention to detail and the clarity of Julia's array abstraction that allowed us to achieve our results without requiring significant amounts of code. We gratefully acknowledge Zak Stone, Michael Isard, Mark Heffernan, James Bradbury, Roy Frostig, Eli Bendersky and Chris Leary of Google's TPU and XLA teams for their openness and willingness to answer our questions about TPUs, answer our bug reports and provide assistance in our quest to make this project a reality. We thank Christopher Rackauckas and Alan Edelman for helpful comments on earlier drafts of this paper.

\bibliography{main}
\bibliographystyle{sysml2019}

\clearpage
\appendix


\begin{figure*}
    \centering
    \begin{tabular}{ r | c | c | c | c | c | c | c | c }
      HLO Instruction Kind & \multicolumn{8}{c}{Instruction count} \\
      &\multicolumn{4}{c}{Forwards} &
      \multicolumn{4}{|c}{Backwards} \\
      &\multicolumn{2}{c}{Unopt} &
      \multicolumn{2}{|c}{Opt} &
      \multicolumn{2}{|c}{Unopt} &
      \multicolumn{2}{|c}{Opt} \\
      &E & T & E & T & E & T & E & T \\

      \hline
parameter & 2 & 56 & 2 & 52 & 2 & 508 & 2 & 183\\
constant & 6 & 24 & 2 & 20 & 31 & 583 & 3 & 93\\
get-tuple-element & 58 & 98 & 58 & 58 & 58 & 181 & 58 & 58\\
add & 0 & 20 & 1 & 20 & 0 & 304 & 1 & 114\\
reshape & 33 & 33 & 4 & 4 & 122 & 122 & 58 & 58\\
map & 22 & 22 & 19 & 19 & 119 & 119 & 38 & 38\\
multiply & 0 & 0 & 0 & 0 & 0 & 256 & 18 & 56\\
tuple & 0 & 20 & 0 & 0 & 21 & 189 & 21 & 21\\
convolution & 16 & 16 & 16 & 16 & 48 & 48 & 47 & 47\\
transpose & 17 & 17 & 1 & 1 & 59 & 59 & 43 & 43\\
broadcast & 20 & 20 & 17 & 17 & 62 & 62 & 19 & 19\\
less-than & 0 & 0 & 0 & 0 & 0 & 108 & 0 & 36\\
maximum & 0 & 23 & 0 & 23 & 0 & 59 & 0 & 23\\
conditional & 0 & 0 & 0 & 0 & 0 & 108 & 0 & 0\\
reduce & 1 & 1 & 1 & 1 & 20 & 20 & 18 & 18\\
select & 0 & 0 & 0 & 0 & 0 & 0 & 0 & 72\\
reverse & 0 & 0 & 0 & 0 & 16 & 16 & 16 & 16\\
dot & 3 & 3 & 3 & 3 & 9 & 9 & 9 & 9\\
reduce-window & 5 & 5 & 5 & 5 & 5 & 5 & 5 & 5\\
select-and-scatter & 0 & 0 & 0 & 0 & 5 & 5 & 5 & 5\\
exponential & 0 & 2 & 1 & 2 & 0 & 5 & 1 & 3\\
less-than-or-equal-to & 0 & 0 & 0 & 0 & 0 & 5 & 0 & 5\\
divide & 0 & 1 & 0 & 1 & 0 & 3 & 0 & 2\\
subtract & 0 & 0 & 0 & 0 & 0 & 1 & 0 & 1\\
\hline\hline
Total & 183 & 361 & 130 & 242 & 577 & 2775 & 362 & 925
      
    \end{tabular}
    \caption{Breakdown of instruction counts of the Metalhead.jl VGG19 forward pass and backwards pass after compilation to XLA. Both unoptimized (after the Julia frontend) and optimized counts (after an XLA optimization pipeline similar to that used by the CPU backend, but without HLO fusion) are shown. For each, the count is further broken down into instructions in the entry computation (E) and instruction counts in all computations (T) }
    \label{tab:all_instructions}
\end{figure*}

\end{document}